\documentclass[twocolumn,showpacs,preprintnumbers,amsmath,amssymb,prb,superscriptaddress]{revtex4}
\usepackage{graphicx}% Include figure files
\usepackage{amsmath}
\usepackage{bm}% bold math
\usepackage{longtable}
\begin{document}

\title{Role of linear and cubic terms for the drift-induced Dresselhaus spin-orbit splitting in a two-dimensional electron gas}

\author{M. Studer}
\affiliation{IBM Research--Zurich, S\"aumerstrasse 4, 8803
R\"uschlikon, Switzerland} \affiliation{Solid State Physics
Laboratory, ETH Zurich, 8093 Zurich, Switzerland}
\author{M. P. Walser}
\affiliation{IBM Research--Zurich, S\"aumerstrasse 4, 8803
R\"uschlikon, Switzerland} \affiliation{Solid State Physics
Laboratory, ETH Zurich, 8093 Zurich, Switzerland}
\author{S. Baer}
\affiliation{Solid State Physics Laboratory, ETH Zurich, 8093
Zurich, Switzerland}
\author{H. Rusterholz}
\affiliation{FIRST Center for Micro- and Nanosciences, ETH Zurich,
8093 Zurich, Switzerland}
\author{S. Sch\"on}
\affiliation{FIRST Center for Micro- and Nanosciences, ETH Zurich,
8093 Zurich, Switzerland}
\author{D. Schuh}
\affiliation{Institut f\"ur Experimentelle und Angewandte Physik,
Universit\"at Regensburg, 93040 Regensburg, Germany}
\author{W. Wegscheider}
\affiliation{Solid State Physics Laboratory, ETH Zurich, 8093
Zurich, Switzerland}
\author{K. Ensslin}
\affiliation{Solid State Physics Laboratory, ETH Zurich, 8093
Zurich, Switzerland}
\author{G. Salis\footnote{To whom correspondence should be addressed: GSA@zurich.ibm.com}}
\affiliation{IBM Research--Zurich, S\"aumerstrasse 4, 8803
R\"uschlikon, Switzerland}
\begin{abstract}
The Dresselhaus spin-orbit interaction (SOI) of a series of
two-dimensional electron gases (2DEGs) hosted in GaAs/AlGaAs and
InGaAs/GaAs (001) quantum wells (QWs) is measured by monitoring the
precession frequency of the spins as a function of an in-plane
electric field. The measured spin-orbit-induced spin-splitting is
linear in the drift velocity, even in the regime where the cubic
Dresselhaus SOI is important. We relate the measured splitting to
the Dresselhaus coupling parameter $\gamma$, the QW confinement, the
Fermi wavenumber $k_F$ and to strain effects. From this, $\gamma$ is
determined quantitatively, including its sign.
\end{abstract}

\maketitle

The spin-orbit interaction (SOI) couples electron spins to the
orbital motion. In semiconductor quantum structures, SOI often
limits the spin lifetime and therefore needs to be minimized when
spins are to be used for processing or storing information. On the
other hand, SOI has the potential to locally control spins by
electrical means.\cite{Datta1990} The reliable manipulation of spins
is crucial for spin-based quantum computing~\cite{Loss1998} and for
spintronic applications~\cite{Wolf2001}. For a two-dimensional
electron gas (2DEG) hosted in a semiconductor with zinc-blende
structure, there are two main sources for SOI: The Rashba
SOI~\cite{Bychkov1984} and the Dresselhaus
SOI~\cite{Dresselhaus1955}. The Dresselhaus SOI is a bulk property
as it arises form the inversion asymmetry of the crystal. The
resulting spin splitting is cubic in the components of the electron
wave vector $\mathbf{k}=(k_x,k_y,k_z)$. In a 2DEG, the quantum
confinement along the growth direction $z$ leads to bound states
with the expectation value $\langle k_z \rangle=0$ and a quantized
value for $\langle k_z^2 \rangle$. This constriction of the orbital
motion modifies the spin splitting in terms of the in-plane momentum
components: In addition to a cubic dependence, there is also a term
proportional to $\langle k_z^2 \rangle$ that is linear in the
components of the in-plane wave vector
$\mathbf{k}=(k_x,k_y)$.\cite{Dyakonov1986} The strength of both
terms is given by the Dresselhaus coupling constant $\gamma$. The
value of $\gamma$ has been measured quantitatively with various
techniques, including Raman scattering,\cite{Jusserand1992} magneto
transport,\cite{Dresselhaus1992,Miller2003,Krich2007} spin-dephasing
measurements~\cite{Marushchak1983} and spin-grating
measurements.\cite{Koralek2009} These techniques probe $\mathbf{k}$
states selectively at the Fermi surface, and the spin splitting is
directly related to the Fermi wave number $k_{\textrm F}$. In
contrast, drift-related
experiments~\cite{Kalevich1990,Kato2004,Meier2007,Wilamowski2007,Studer2009}
involve a small displacement $\delta \mathbf k$ of the Fermi surface
and lead to a drift-induced spin splitting that is proportional to
$\delta k \ll k_{\textrm F}$ and a direction dependence that can be
described by an effective magnetic field
$\mathbf{B}_\textrm{df}(\delta \mathbf k)$. This field is composed
of a Rashba component $\mathbf{B}_\textrm{df,R}$ and a Dresselhaus
component $\mathbf{B}_\textrm{df,D}$, which have different symmetry
with respect to $\delta \mathbf k$. Quantitative values, including
the absolute sign, for both Rashba and Dresselhaus SOI can be
obtained by monitoring the coherent spin precession in an external
magnetic field
$\mathbf{B}_\textrm{ext}$.\cite{Kato2004,Meier2007,Studer2009}

In this work, we investigate the interplay of the linear and cubic
terms of the Dresselhaus SOI of a 2DEG as manifested in measurements
of $\mathbf{B_\textrm{df,D}}$ versus $\mathbf{\delta k}$. The cubic
term becomes important for $k_F^2>\langle k_z^2\rangle$, and it has
been suggested that in this regime $\mathbf{B}_\textrm{df}$ has a
cubic dependence on $\mathbf{\delta
k}$.\cite{Meier2007,Chernyshov2009,Norman2010,Bernevig2005} Here we
show that $\mathbf{B}_\textrm{df,D}$ depends linearly on $\delta k$
($\ll k_\textrm F$), even in the regime of $k_\textrm F^2 \gg
\langle k_z^2\rangle$. A central aspect of this work is to discuss
how the linear and the cubic Dresselhaus SOI terms contribute to
$\mathbf{B}_\textrm{df,D}$ by deriving an explicit expression for
$\mathbf{B}_\textrm{df,D}(\mathbf{\delta k})$ from the Dresselhaus
Hamiltonian $H_D$. The calculations show that the slope of
$B_\textrm{df,D}$ versus $\delta k$ is decreased by the cubic term,
and changes its sign when $k_\textrm F^2 > 2\langle k_z^2\rangle$.
We experimentally studied samples made of two material systems
displaying different Fermi and confinement energies, namely 10-nm
and 20-nm-wide InGaAs/GaAs QWs, and 15-nm-wide GaAs/AlGaAs QWs. A
linear dependence of $B_\textrm{df,D}$ on $\delta k$ is measured in
all cases of $\langle k_z^2 \rangle/k_\textrm F^2$. Cubic
contributions are small for the GaAs/AlGaAs QWs and we find $\gamma
\approx -6$~eV\AA$^3$. For the InGaAs/GaAs QWs, cubic contributions
significantly modify the slope of $B_\textrm{df,D}$ versus $\delta
k$. In addition, a more detailed analysis of the slope provides
evidence of a strain-induced contribution that has the same symmetry
as the linear Dresselhaus SOI.

The paper is organized as follows: First we lay out the theoretical
framework and define the Dresselhaus SOI and the coordinate system,
the basis for an unambiguous definition of the sign of $\gamma$.
Then, we derive the linear dependence of $B_\textrm{df}$ versus
$\delta k$. Next, we present experimental data for a series of
samples in which the importance of the cubic term varies, and
finally discuss the results with respect to the theory developed.

\section{Dresselhaus spin-splitting}

Dresselhaus SOI for a 2DEG is described here in the coordinate
system ($x,y,z$) defined by the cubic crystallographic axes $[100],
[010], [001]$. These directions are specified in a GaAs primitive
cell, with the cation (Ga) residing at (0, 0, 0) and the anion (As)
located at ($\frac{1}{4}$, $\frac{1}{4}$, $\frac{1}{4}$). This is
the same convention as used by wafer manufacturers, and is
compatible with the literature on direction-selective etching
processes.\cite{Shaw1981,Adachi1983} In contrast, in some
theoretical works,\cite{Cardona1986,Cardona1988} the anion is placed
at the origin of the coordinate system. This specification is
important for an unambiguous definition of the Dresselhaus parameter
$\gamma$ because the exchange of cation and anion involves a change
in the sign of $\gamma$.\cite{Cardona1986} In the following, we
discuss (001)-grown QWs with one occupied subband and the
Dresselhaus Hamiltionian
\begin{equation}
 H_D = \gamma \bigg[-\langle k_z^2 \rangle(k_x \sigma_x - k_y \sigma_y)+ (k_y^2 k_x \sigma_x  - k_x^2 k_y \sigma_y) \bigg],
%\nonumber \\
\label{eq:BIAham}
\end{equation}
where $\langle k_z^2 \rangle$ is the quantum-mechanical expectation
value of $k_z^2$ with respect to the envelope wave function of the
occupied subband, and $\vec{\sigma}=(\sigma_x,\sigma_y,\sigma_z)$
are the Pauli matrices. In this notation, two terms show up: A term
that is linear in the in-plane momentum components and a term that
is cubic. To study the interplay of these two terms, we compare
their magnitude for $\mathbf{k}=(k_x,k_y)$ on the Fermi surface,
i.e. for $k_x^2+k_y^2=k_F^2$: The ratios $k_x^2/\langle k_z^2
\rangle$ and $k_y^2/\langle k_z^2 \rangle$ have an upper limit of
$k_F^2/\langle k_z^2 \rangle$ and are zero for certain orientations
of $\mathbf{k}$. Therefore, the linear term is dominant in the case
of strong confinement, i.e. for $\langle k_z^2 \rangle \gg k_F^2$.
In terms of the QW sheet density $n_s$ and the QW width $w$, the
importance of cubic terms scales with the product $n_s w^2$, because
$k_F^2=2\pi n_s$ and $\langle k_z^2 \rangle\propto(\pi/w)^2$.

For the further discussion, it is convenient to express $H_D$ as a
spin-splitting induced by an effective magnetic field
$\mathbf{B}_\textrm{D}$ that depends on the electron wave vector
$\mathbf{k}$:

\begin{equation}
E(\mathbf{k},\uparrow)-E(\mathbf{k},\downarrow)= g\mu_{B}|\mathbf{B}_{D}(\mathbf{k})|.
\label{eq:Zeeman}
\end{equation}

Here $\mu_{B}=|e|\hbar/2m_0$  is Bohr's magneton of the free
electron with mass $m_0$, and $g$ is the effective $g$-factor in the
semiconductor QW. $E(\mathbf{k},\uparrow)$ and
$E(\mathbf{k},\downarrow)$ are the two eigenvalues of $H_D$ for a
given $\mathbf{k}$. In our coordinate system and in correspondence
with Eq.~(\ref{eq:BIAham}), we separate the linear and cubic
contributions according to

\begin{equation}\label{eq:BBIA}
\mathbf{B}_\textrm{D}(k_x,k_y)=\mathbf{B}_{{\textrm D},1}(k_x,k_y) + \mathbf{B}_{{\textrm D},3}(k_x,k_y)
\end{equation}

\begin{equation}\label{eq:BBIA1}
\mathbf{B}_{{\textrm D},1}(k_x,k_y)=\frac{-2\gamma
\langle k_z^2 \rangle}{g \mu_B} \binom{k_x} {-k_y}
\end{equation}

\begin{equation}\label{eq:BBIA2}
\mathbf{B}_{{\textrm D},3}(k_x,k_y)=\frac{2 \gamma}{g \mu_B} \binom{k_xk_y^2}{-k_yk_x^2}.
\end{equation}

\begin{figure}[htb]
\includegraphics[width=84mm]{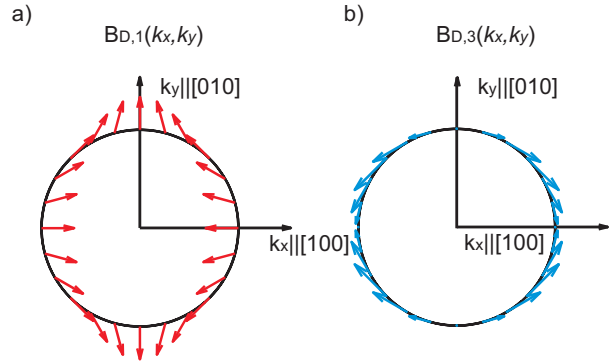}
\caption{\label{fig:fig1} (Color online) Symmetry of the effective
magnetic field resulting from Dresselhaus SOI: (a) linear term
$\mathbf{B}_{D,1}$  and (b) cubic term $\mathbf{B}_{D,3}$ (we assume
$g<0$ and $\gamma<0$).}
\end{figure}

The different symmetries of the two contributions are illustrated in
Fig.~\ref{fig:fig1} by plotting the vector fields
$\mathbf{B}_{D,1}(k_x,k_y)$ and $\mathbf{B}_{D,3}(k_x,k_y)$ for a
specific $|\mathbf{k}|$. The magnitude of $\mathbf{B}_{D,1}$ does
not depend on the direction of $\mathbf{k}$, whereas
$\mathbf{B}_{D,3}$ is maximal for $\mathbf{k}$ along the $[110]$ and
$[1\overline{1}0]$ direction, and zero along $[100]$ and $[010]$.

So far, we have only considered Dresselhaus SOI. Other sources of
SOI include structure inversion asymmetry (Rashba SOI) and strain.
Their corresponding spin splitting can also be expressed as an
effective magnetic field. In case of the Rashba SOI, this field can
be separated from the Dresselhaus SOI because of its different
symmetry in $\mathbf{k}$. However, biaxial strain gives rise to a
spin splitting that has the same symmetry as the linear Dresselhaus
SOI [first term in Eq.~(\ref{eq:BIAham})]. Its effective magnetic
field can be expressed as\cite{Meier1984,Bernevig2005}

\begin{equation}\label{eq:BBIAS}
\boldsymbol{B}_{\textrm{S}}(k_x,k_y) = \frac{2\beta_S}{g \mu_B}
\binom{k_x}{-k_y}.
\end{equation}
Here, $\beta_S=D(\epsilon_{zz}-\epsilon_{xx})$, where $D$ is a
material parameter and $\epsilon_{ij}$ is the strain tensor. Such a
component has been observed in InGaAs epilayers grown on GaAs, where
it can be stronger than the Dresselhaus contribution.\cite{Kato2004}
To our knowledge, there is no systematic work describing the strain
contribution in $n$-doped QWs and only few experimental
studies.\cite{Studenikin2003} For completeness and
cross-referencing, we introduce the Rashba coupling constant
$\alpha$ and the effective magnetic field induced by Rashba SOI:

\begin{equation}
\textbf{B}_\textrm{R}=\frac{2\alpha}{g \mu_B}\binom{k_y}{-k_x}.
\end{equation}

\section{Calculation of the drift-induced spin splitting}

\begin{figure}[htb]
\includegraphics[width=84mm]{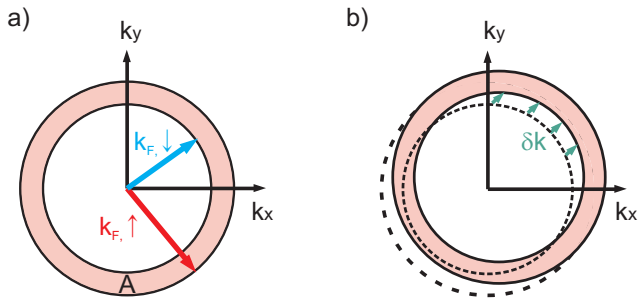}
\caption{\label{fig:fig2} (Color online) (a) 2DEG  with finite spin
polarization: Two Fermi circles with radii $k_{F,\uparrow}$ and
$k_{F,\downarrow}$ limit the spin-polarized $k$-space area
$A=\pi(k_{F,\uparrow}^2-k_{F,\downarrow}^2)$ [shaded (red)]. (b)
Spin-polarized 2DEG in an in-plane electric field: The Fermi circles
and the area A are shifted by a drift wave vector $\delta
\textbf{k}$}.
\end{figure}

The effective magnetic field $\mathbf{B}_\textrm{D}(\mathbf{k})$
discussed in the previous section affects individual electron spins.
Drift-related experiments involve all states near the Fermi surface,
and a corresponding average of $\mathbf{B}_\textrm{D}(\mathbf{k})$
is observed. In our experiment, only a part of the 2DEG is spin
polarized, and we measure the coherent precession of those spins in
their corresponding field $\mathbf{B}_\textrm{D}(\mathbf{k})$. The
initial spin polarization is oriented along $\mathbf{z}\equiv[001]$
(denoted as "spin up" polarization), and can be described by two
Fermi wave numbers $k_{F,\uparrow}$ and $k_{F,\downarrow}$: Spin-up
(spin-down) states are homogeneously filled up to $k_{F,\uparrow}$
($k_{F,\downarrow}$) at 0~K [Fig.~\ref{fig:fig2}(a)]. The spin
polarization is given as $(k_{F,\uparrow}^2 -k_{F,\downarrow}^2)/2
\pi n_s $.

To connect $\mathbf{B}_\textrm{D}(\mathbf{k})$ with the
spin-precession frequency measured in the experiment, we average
$\mathbf{B}_\textrm{D}(\mathbf{k})$ over the spin-polarized area
$A=\pi(k_{F,\uparrow}^2-k_{F,\downarrow}^2)$ [shaded gray (red) in
Fig.~\ref{fig:fig2}(a)]. This approach is reasonable for
nonselective optical probe techniques and for fast electron momentum
scattering. In equilibrium, the average over $A$ is obviously zero
for the rotationally symmetric fields
$\mathbf{B}_\textrm{D,1}(\mathbf{k})$ and
$\mathbf{B}_\textrm{D,3}(\mathbf{k})$. However, if the 2DEG is
exposed to an in-plane electric field $\mathbf{E}$,  the
spin-polarized states are shifted from the equilibrium position in
the direction of the drift wave vector $\mathbf{\delta  k}=(\delta
k_x,\delta k_y)=-\mu\mathbf{E} m^* /\hbar$ [Fig.~\ref{fig:fig2}(b)].
Note that $\mathbf{E}$ points in the opposite direction than
$\mathbf{\delta k}$ because the electron mobility $\mu$ is positive
by definition and $\delta \mathbf{k}$ corresponds to electrons with
a negative charge. Using polar coordinates $\mathbf{k}=(r
\cos(\xi),r\sin(\xi))$, the average over the shifted area $A$ is
given by the integral

\begin{widetext}
\begin{eqnarray}\label{eq:integral}
\textbf{B}_\textrm{df,D}(\mathbf{\delta  k})
 = \frac{1}{A} \int_{k_{F,\downarrow}}^{k_{F,\uparrow}}\!\!\!\!\!\!\!\!r dr \int_{0}^{2 \pi}\!\!\!\!\!d\xi
 \,\,[\mathbf{B}_{D}(\mathbf{k}+\delta \mathbf{k})+\mathbf{B}_{S}(\mathbf{k}+\delta \mathbf{k})]
=\frac{2 \gamma [-\langle k_z^2
\rangle\!+\!\frac{1}{4}(k_{F,\downarrow}^2+k_{F,\uparrow}^2)]+2\beta_S}{g
\mu_B} \binom{\delta k_x} {-\delta k_y} +\mathcal{O}(\delta k^3).
\label{eq:integralResults}
\end{eqnarray}
\end{widetext}

This averaging transforms $\mathbf{B}_{D}(\mathbf{k})$ and
$\mathbf{B}_{S}(\mathbf{k})$ into a drift field
$\textbf{B}_\textrm{df,D}(\mathbf{\delta  k})$ that expresses the
average field experienced by the polarized electron spins. The
dominant term in $\mathbf{B}_\textrm{df,D}$ is linear in
$\mathbf{\delta k}$, and third-order terms in $\mathbf{\delta k}$
are unimportant for $\delta k^2 \ll \langle k_z^2 \rangle$ and
$\delta k^2 \ll k_{F}^2$. We note that the terms proportional to
$k_{F,\uparrow}^2$ and $k_{F,\downarrow}^2$ originate from
$\mathbf{B}_{D,3}$, i.e., the cubic Dresselhaus term affects
$\mathbf{B}_\textrm{df,D}$ in linear order. The symmetry of
$\mathbf{B}_\textrm{df,D}$ is the same as that of
$\mathbf{B}_\textrm{D,1}$, and the cubic terms
$\mathbf{B}_\textrm{D,3}$ do change neither this symmetry nor the
linearity in $\delta \mathbf{k}$. For small spin polarization, we
can approximate $k_{F,\uparrow}$ and $k_{F,\downarrow}$ in
Eq.~(\ref{eq:integralResults}) by the Fermi wave number $k_{F}$ of
the unpolarized spins.  In a previous publication the influence of
the cubic Dresselhaus term has been underestimated by a factor of
two because of the assumption of a high spin
polarization.\cite{Studer2009} Using the above described
simplification, reasonable for a typical experiment, we define the
drift-field Dresselhaus coefficient

\begin{equation}
\beta^*:=\gamma(-\langle k_z^2 \rangle + \frac{1}{2}k_F^2)+\beta_S,
\label{eq:betastar}
\end{equation}
 We rewrite
the dominant term of Eq.~(\ref{eq:integralResults})
\begin{equation}
\mathbf{B}_\textrm{df,D}(\delta k_x,\delta k_y) = \frac{2\beta^*}{g
\mu_B} \binom{\delta k_x}{-\delta k_y}.
\label{eq:BBIAD}
\end{equation}

The importance of Eq.~(\ref{eq:BBIAD}) is based on the direct
experimental accessibility of $\beta^*$, whereas the parameter
$\beta=-\gamma \langle k_z^2 \rangle$ has to be calculated from
$\beta^*$, $\langle k_z^2 \rangle$, $k_{\textrm F}^2$ and $\beta_S$.
Temperature broadening leads to an increase of the cubic Dresselhaus
contributions in Eq.~\ref{eq:betastar} that becomes significant if
$E_F$ is comparable to $k_B T$.\cite{Weng2004} In our measurements
this correction is below 10~\%.

\section{Measurement of the Dresselhaus SOI}

We investigate Si-doped $n$-type material systems, namely
epitaxially grown GaAs/AlGaAs QWs and InGaAs/GaAs QWs. All QWs have
been grown along [001]. The InGaAs/GaAs QWs are single QWs, whereas
the GaAs/AlGaAs samples consist of a series of 14 equivalent QWs.
Narrow GaAs/AlGaAs QWs are suited for a reliable determination of
the Dresselhaus coupling constant $\gamma$, because GaAs/AlGaAs is a
nearly strain-free system, and cubic contribution are small for
narrow QWs. InGaAs/GaAs QWs with less confinement and large electron
density are used to study cubic Dresselhaus SOI and to test the
consistency with a strain-induced contribution.
Table~\ref{tab:Samples} summarizes the parameters of all QWs that
were used.

All optical measurements have been done at elevated temperature of
40~K, to preclude nuclear spin effects. The sheet carrier density
$n_s$ of the 2DEGs was measured separately in a perpendicular
magnetic field, using both Hall and Shubnikov--de Haas measurements.
Both methods yield equivalent $n_s$  at 1.6\,K. The mobility was
then calculated from the four-terminal resistance. A quantity that
can not be accessed in the experiment is $\langle k_z^2 \rangle$. In
previous publications, $\langle k_z^2 \rangle\approx(\pi/w)^2$ was
used, where $w$ is the width of the QW.\cite{Studer2009,Koralek2009}
This approximation is based on a square potential QW with an
infinitely high confinement potential in the $z$ direction. Note
that this is an appropriate estimate only for QWs with a very large
conduction-band offset. To improve the accuracy of the
experimentally determined value for $\gamma$, we have used a
one-dimensional Poisson and Schr\"odinger equation solver
\footnote[0]{Calculations were done with the Nextnano$^3$ software
version 2004-Aug-24} to calculate the envelope wave function of the
ground state $\psi(z)$, and then numerically determined the
expectation value $\langle k_z^2 \rangle:=\int [\psi'(z)]^2 dz$. The
numerical approach yields $\langle k_z^2 \rangle
=2.5\!\times\!10^{16}$~m$^{-2}$ for a 15-nm-broad
GaAs/Al$_{.3}$Ga$_{.7}$As QW. This is considerably less than for an
infinite barrier [$(\pi/15~nm)^2=4.4\!\times\!10^{16}$~m$^{-2}$],
because of leakage of the wave function into the barrier region and
electron screening.

\begin{table*}
\caption{\label{tab:Samples} Overview of all samples:
      QW material, structure, width $w$, and expected $\langle k_z^2 \rangle$ (calculated);
      $n_s$ and $k_F^2$ are experimental results from Hall measurements, and $\mu$ is
      deduced from the four-terminal resistance. The Dresselhaus drift parameter $\beta^*$
      is determined from the drift-induced change of the spin-precession frequency, using TRFR
      (Sample 1 and 2) and TRKR (Samples 3-5). $\gamma$ is calculated from $\beta^*$, $\langle k_z^2 \rangle$,
      and $k_F^2$. For sample 1 and 2 the calculation of $\gamma$ in addition assumes the same
      $\beta_S\approx 7\!\times\!10^{-14}\,$eVm. $\beta$ is the
      linear term of the Dresselhaus SOI and calculated using $\beta=-\gamma\langle k_z^2 \rangle$. All data in this table has been obtained at a temperature of 40 K.}

    \begin{tabular}{|c|c|c|c|c|c|c|c|c|c|}
        \hline
          & QW/host material   &     $w$     & $\mu$   &$\langle k_z^2 \rangle \!\times\! 10^{16}$ &$k_F^2 \!\times\! 10^{16}$     &  $n_s\!\times\! 10^{15}$  & $\beta^*\!\times\! 10^{-13}$  & $\gamma\!\times\!10^{-30}$ & $\beta\!\times\!10^{-13}$\\
        no.     &  (number of wells)  &     (nm)      &  (m$^2$/Vs)        &     (m$^{-2})$                        &             (m$^{-2}$)    & (m$^{-2}$)         &    (eVm)   &    (eVm$^3$)  &    (eVm)  \\
        \hline
        1 & In$_{.10}$Ga$_{.90}$As/GaAs (1~$\!\times$) & 20    & 1.3        &0.95                                    & 3.2                       &  5.2                  &    0.2    & -7.5&    0.7   \\
        2 & In$_{.12}$Ga$_{.88}$As/GaAs (1~$\!\times$) & 10    & 4.5        &2.25                                    & 2.8                       &  4.5                  &    1.3 & -7.5 &   1.7    \\
        3 & GaAs/Al$_{.3}$Ga$_{.7}$As (14~$\!\times$)   & 15    & 51         &2.5                                     & 0.74                      &  1.2                  &    1.7   & -8.0&    2.0   \\
        4 & GaAs/Al$_{.3}$Ga$_{.7}$As (14~$\!\times$)   & 15    & 36         &2.5                                     & 0.86                      &  1.4                 &     1.0  & -4.7&    1.2   \\
        5 & GaAs/Al$_{.3}$Ga$_{.7}$As (14~$\!\times$)   & 15    & 24         &2.5                                     & 0.99                      &  1.6                 &     1.0  & -5&    1.3   \\

        \hline
      \end{tabular}

\end{table*}

To apply an in-plane electric drift field $\mathbf{E}$ in a specific
direction, a mesa structure is etched into the substrates hosting
the 2DEG to form 100-$\mu$m or 150-$\mu$m-wide conductive channels.
The geometry of the mesa structure is cross-shaped for samples 1 and
3, whereas unidirectional mesa bars were used for the other samples.
The electric field is inhomogeneous in the center of the
cross-shaped mesa, a fact that has been accounted for in the
evaluation.\cite{Studer2009} Ohmic contacts to the buried channel
are subsequently fabricated by standard AuGe diffusion. In the
experiment, a small spin-polarization of the electron spins in the
conduction band of the QW is generated by a circularly polarized
pump-pulse generated by a mode-locked Ti:sapphire laser. The energy
per area of the pump pulses was kept below
2$\times$10$^{-2}$\,Jm$^{-2}$ for sample 1 and 2, and below
4$\times$10$^{-3}$\,Jm$^{-2}$ for samples 3-5. These small
intensities ensure that the number of electron-hole pairs that are
excited per QW is much smaller than the number of electrons already
in the QW. Therefore, only a small fraction of the electrons become
spin-polarized by the pump pulse.

The experiment then monitors, in the time domain, the coherent
precession of electron spins in an external magnetic field $\mathbf
B_\textrm{ext}$ using time-resolved Kerr rotation (TRKR) or
time-resolved Faraday rotation (TRFR).\cite{Meier2007} The spin
precession frequency is given by

\begin{equation}
\Omega(\mathbf{E}) = \frac{|g| \mu_B}{\hbar}\,
|\mathbf{B}_\textrm{ext}+\mathbf{B}_\textrm{df}(\mathbf{E})|.
\label{eq:omega}
\end{equation}

$\mathbf{B}_\textrm{ext}$ is applied in the plane of the QW, and
SOI-induced contributions to the spin splitting become visible as an
$\mathbf{E}$-field-dependent change of $\Omega$ given by $\Delta
\Omega(\mathbf{E})=\Omega(\mathbf{E})-\Omega(0)$. The observation of
several periods of spin precession during the spin lifetime is
required to precisely determine $\Omega$ and thus $\Delta \Omega$.
This is made possible by exposing the sample to $\mathbf
B_\textrm{ext}\approx1\,\textrm{T}$. In bulk GaAs, where the spin
lifetime is much longer than in QWs, coherent spin precession about
$\mathbf{B}_\textrm{df}$ has been observed at $\mathbf
B_\textrm{ext}=0\,\textrm{T}$.\cite{Kato2004}

For our experiments, $|\mathbf{B}_\textrm{df}| \ll
|\mathbf{B}_\textrm{ext}|$, and therefore $\Delta \Omega$ can be
approximated by

\begin{equation}\label{eq:omegaProj}
\Delta \Omega(\mathbf{E})
\approx \frac{|g|\mu_B}{\hbar}  \frac{\mathbf{B}_\textrm{df}(\mathbf{E})\cdot \mathbf{B}_\textrm{ext}}{|\mathbf{B}_\textrm{ext}|},
\end{equation}
i.e. $\Delta \Omega$ is proportional to the projection of
$\mathbf{B}_\textrm{df}$ along the direction of the external field.
By selecting the orientation of $\mathbf{B}_\textrm{ext}$ with
respect to $\mathbf{E}$, the contributions with Dresselhaus
symmetry, $\Delta \Omega_D$, can be separated from contributions
having another symmetry. We used two different methods for this
separation:

\begin{eqnarray}
\textrm{(I)}\,\,\,\, \mathbf{E} \perp \mathbf{B}_\textrm{ext}
&\textrm{, with }& \mathbf{B}_\textrm{ext}\parallel[ 1\overline{1}0] \textrm{\,\,or\,\,} [110]\nonumber \\
\textrm{(II)}\,\,\,\mathbf{E} \parallel \mathbf{B}_\textrm{ext}
&\textrm{, with }& \mathbf{B}_\textrm{ext}\parallel [100] \textrm{\,\,or\,\,} [010].\nonumber
\end{eqnarray}

For (I), $\Delta \Omega(\mathbf{E})$ is recorded for two
orientations of
$\mathbf{B}_\textrm{ext}$,~\footnote[2]{$\mathbf{B}_\textrm{df,D}$
and $\mathbf{B}_\textrm{df,R}$ are either parallel or antiparallel
for this set of directions.} and the sum  of these two measurements
yields the Dresselhaus contribution $\Delta
\Omega_D$.\cite{Meier2007} For (II), $\Delta \Omega$ is a direct
measure of the Dresselhaus contribution $\Delta
\Omega_D$,\cite{Meier2007}
~\footnote[3]{$\mathbf{B}_\textrm{df,D}\perp
\mathbf{B}_\textrm{df,R}$ for this set of directions, and
$\mathbf{B}_\textrm{df}\cdot
\mathbf{B}_\textrm{ext}=\mathbf{B}_\textrm{df,D}\cdot
\mathbf{B}_\textrm{ext}$.} i.e.,

\begin{equation}\label{eq:omegaBIA}
\Delta \Omega (\mathbf{E})=\Delta \Omega_D (\mathbf{E}) \approx \pm
\frac{|g|\mu_B}{\hbar} {|\mathbf{B}_\textrm{df,D}(\mathbf{E})|}.
\end{equation}

\begin{figure}[htb]
\includegraphics[width=84mm]{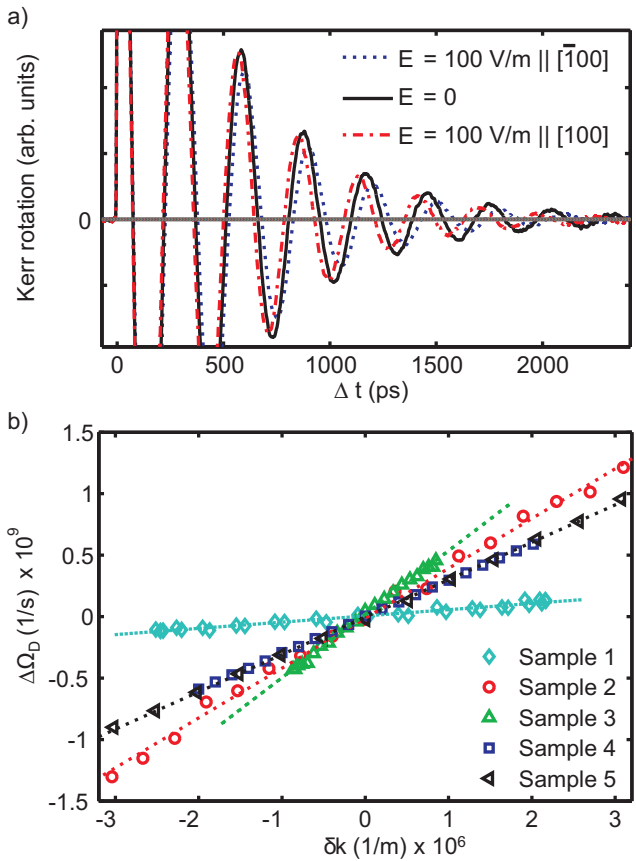}
\caption{\label{fig:fig3} (Color online) (a) TRKR measurements as a
function of the electric drift-field $E$ (Sample 4,
$\mathbf{B}_\textrm{ext}$ along $[100]$,
$|\mathbf{B}_\textrm{ext}|$=0.93~T). (b) The Dresselhaus
drift-induced frequency-shift $\Delta\Omega_D$ depends linearly on
the drift wave-vector $\delta k$ for all samples measured.}
\end{figure}

As an example, data from sample 4 is shown where we used method II
to directly see the influence of the Dresselhaus SOI.
Figure~\ref{fig:fig3}(a) shows experimental TRKR traces measured
with $\mathbf{B}_\textrm{ext}$=0.93~T applied along $[100]$. The $g$
factor $|g|$ is obtained from the trace at $E=0$. Also the data for
 $\mathbf E$ along $[\overline{1}00]$ and  $\mathbf
E$ along $[100]$ have been plotted. In this configuration, $\Omega$
increases for $\mathbf{E}$ along $[100]$.

As a central finding, we show that $\Delta\Omega_D$ depends linearly
on the drift wave-number $\delta k$ [Fig.~\ref{fig:fig3}(b)]. This
is supported by experimental data from various samples (listed in
Table~\ref{tab:Samples}). For sample 1 and 2, $\Delta\Omega_D$ has
been obtained using method (I), and for sample 3, 4 and 5, we used
method (II). According to Eqs.~(\ref{eq:BBIAD}) and
(\ref{eq:omegaBIA}), the slope of the fit in Fig.~\ref{fig:fig3}(b)
is given by $2\beta^*/\hbar$. The sign of $\beta^*$ is determined
from the direction of $\mathbf{B}_\textrm{df,D}$ with respect to
$\mathbf{B}_\textrm{ext}$: For example, for $\mathbf{E}$ along
$[\overline{1}00]$ ($=-\hat{x}$), $\delta \mathbf{k}$ points along
$[100]$ and, according to Eq.~(\ref{eq:BBIAD}),
$\mathbf{B}_\textrm{df,D}$ points along $[\overline{1}00]$ for
$\beta^*>0$ and $g<0$. From Eq.~(\ref{eq:omegaProj}), we see that
$\Delta \Omega_D$ is negative if $\mathbf{B}_\textrm{ext}$ is
applied along $[100]$. Therefore, the sign of $\beta^*$ must be
positive for the data shown in Fig.~\ref{fig:fig3}(a). As seen from
Table\,\ref{tab:Samples}, we find $\beta^*>0$ for all samples
investigated.

\section{Discussion}
First, we will discuss the results for GaAs/AlGaAs and compare
$|\gamma|$ with the literature values. Second, we present the
results for InGaAs/GaAs, for which we expect a significant
contribution to SOI from third-order terms and strain. Finally, we
compare the sign of $\gamma$ we obtained with the results of other
experimental work.

As GaAs/AlGaAs grows nearly strain-free, we assume $\beta_S=0$ and
calculate $\gamma$ directly from the measured $\beta^*$ using
Eq.~(\ref{eq:betastar}) and the values of $\langle k_z^2 \rangle$
and $k_F^2$ (Table~\ref{tab:Samples}). For $\gamma$, we find values
in the range of -4 to -8~eV\AA$^3$. These values are smaller than
previously reported for GaAs/AlGaAs
QWs~\cite{Jusserand1995,Jusserand1992,Dresselhaus1992,Miller2003}
and GaAs bulk crystals,\cite{Marushchak1983} where $|\gamma|$ covers
a rather wide range of 11-27~eV\AA$^3$. Recent results from
spin-grating measurements yield $|\gamma|=5$~eV\AA$^3$, assuming
$\langle
k_z^2\rangle=(\pi/w)^2=6.8\!\times\!10^{16}\,$m$^{-2}$.\cite{Koralek2009}
For these QWs, our simulations suggest $\langle k_z^2
\rangle=3.4\!\times\!10^{16}\,$m$^{-2}$, which corresponds to
$|\gamma|=10\,$eV\AA$^3$. A similar result of
$|\gamma|\approx8$~eV\AA$^3$ was found for a quantum dot hosted in
an AlGaAs/GaAs heterostructure.\cite{Krich2007} Measurements of
spin-orbit-induced spin precession at
$B_\textrm{ext}=0$\cite{Leyland2007} and of $g$ factor
anisotropy\cite{Eldridge2010B} yield values of $|\gamma|$ between 4
and 10~eV\AA$^3$.

Results from theoretical calculations are in the range of
$|\gamma|=8-36\,$eV\AA$^3$ (see supplementary notes of
Ref.~\onlinecite{Krich2007}).

There are several possibilities for an error in the experimental
value for $\gamma$. In particular, a systematic underestimation of
$\gamma$ could result from a reduction of the drift wave-number
$\delta k$ under optical illumination. The electric field $E$ and
therefore $\delta k$ have been determined from the four-terminal
resistance measured simultaneously with the TRKR (TRFR). The pump
pulse locally creates electron-hole pairs that initially screen the
applied electric field. This effect can be reduced by using pump
pulses with smaller intensities. Indeed, we observe a small
dependence of $\beta^*$ upon the pump power. For sample 5, $\beta^*$
increases to about 1.2$\times$10$^{-13}$\,eVm if the pump intensity
is decreased by a factor of 8 (reaching
5$\times$10$^{-4}$Jm$^{-2}$). We have not observed a persistent
photoeffect: The average electron density as determined from both
Hall and Shubnikov--de Haas measurements did not change
significantly upon illumination with a light-emitting diode.

To investigate the effect of cubic Dresselhaus SOI and to quantify
the importance of cubic terms compared with that of linear terms, we
have chosen samples with large $n_s$ and small $\langle k_z^2
\rangle$, such that $k_F^2 \gtrsim 2\langle k_z^2 \rangle$ (sample
1). For this sample, we find $\beta^*>0$, which at first sight is
surprising because $\beta^*$ is expected to change sign for $k_F^2 >
2\langle k_z^2 \rangle$ if $\beta_S=0$ [Eq.~(\ref{eq:betastar})].
The positive sign can be explained if strain affects the SOI. For
simplicity, we assume that samples 1 and 2 have the same $\gamma$
and $\beta_S$. With the values of $\beta^*,\langle k_z^2\rangle$ and
$k_F^2$ listed in table~\ref{tab:Samples}, we solve the set of
equations given by Eq.~(\ref{eq:betastar}) with the two unknowns
$\beta_\textrm S$ and $\gamma$. As a result we obtain
$\beta_S\approx7\!\times\! 10^{-14}$~eVm and
$\gamma\approx-7.5\,$eV\AA$^3$. This value for $\gamma$ is in rather
good agreement with the results for the GaAs/AlGaAs QWs and suggests
that 2D structures can exhibit strain-induced spin-splitting with
magnitudes comparable to the linear Dresselhaus SOI
($\beta_S\approx\beta\approx7\!\times\! 10^{-14}$~eVm for sample 1).
In this analysis, the biaxial strain component directly affects the
resulting $\beta^*$, and further more systematic investigations are
needed to obtain a more precise value for $\gamma$ in InGaAs/GaAs
QWs.

In our measurements, the sign of $\gamma$ is  negative for all
samples. Some care has to be taken to determine the sign of
$\gamma$. We have experimentally verified the sign of the magnetic
field in our lab. Moreover, the sign is based on the assumption that
$g<0$ and that the drift direction is the opposite direction of
$\mathbf{E}$, i.e. $\delta \mathbf{k}\parallel - \mathbf{E}$. We
have kept track of the crystallographic direction indicated by the
wafer manufacturer during processing and double checked with a
selective etching test.\cite{Shaw1981,Adachi1983} The often cited
positive sign of $\gamma$ is specified in an As-based coordinate
system, which translates to a negative sign in a Ga-based
system.\cite{Cardona1986}

In Ref.~\cite{Riechert1984} the sign of $\gamma$ for a GaAs bulk
crystal has been measured by analyzing the phase that an electron
spin acquires during traveling. A positive $\gamma$ has been
obtained in an As-based coordinate system, in agreement with our
result. The sign of $\gamma$ can also be compared with experiments
that only determine the sign of the ratio $\alpha/\beta$ if the sign
of $\alpha$ is known. From previous measurements on a parabolic QW
with back and front gates,\cite{Studer2009PRL} the sign of $\alpha$
has been determined to be positive for an electric field
$\mathbf{E_{\perp}}$ pointing along the growth direction
$(+\mathbf{z})\parallel[001]$. From $\beta=-\gamma \langle k_z^2
\rangle>0$, it follows that $\alpha/\beta$ is positive for
$\mathbf{E_{\perp}}\parallel(+\mathbf{z})$. This is in agreement
with Ref.~\onlinecite{Frolov2009}: Using a ballistic spin resonance
experiment, a negative sign of $\alpha/\beta$ has been determined in
the heterostructure investigated with $\mathbf{E_{\perp}}$ along
$-\mathbf{\hat{z}}$. It is also compatible with
Ref.~\onlinecite{Larionov2008}: $\alpha/\beta$ is determined from
the spin-dephasing anisotropy and becomes negative for  bias
voltages larger than 1.2~V. In that experiment, $\beta$ is defined
as $+\gamma \langle k_z^2 \rangle$ ($=-\beta$ in our definition),
and a positive bias means a negative voltage on the front gate and
ground potential on the back gate, i.e., $\mathbf{E_{\perp}}$ is
along $+\mathbf{\hat{z}}$~\footnote[1]{L. E. Golub (private
comunication)}.

To summarize, we have shown how SOI in 2DEGs manifests itself in
drift-related experiments: The drift-induced spin-splitting due to
Dresselhaus SOI is linear in the drift velocity given by the drift
wave-number. The linear proportionality is characterized by
$\beta^*$. Cubic Dresselhaus SOI becomes important for a large
electron density and weak electron confinement, and lowers the value
of $\beta^*$ by a factor of $1-\frac{1}{2} k_\textrm{F}^2/\langle
k_z^2 \rangle$. Already in a 15-nm-wide GaAs/Al$_{.3}$Ga$_{.7}$As QW
with an electron density of 1.6$\times$10$^{15}$\,m$^{-2}$,
$\beta^*$ is reduced by 20\% compared with $\beta=-\gamma \langle
k_z^2 \rangle$. We have experimentally determined $\beta^*$ from the
drift-induced change in the spin-precession frequency. The
Dresselhaus coupling constant $\gamma$ was obtained from the
$\beta^*$ measured, using the electron density from Hall
measurements and the simulation result for $\langle k_z^2 \rangle$.
For GaAs/AlGaAs QWs, we find $\gamma\approx -6\,$eV\AA$^3$, and for
an InGaAs/GaAs QWs, $\gamma\approx-8~$eV\AA$^3$. The results for
InGaAs/GaAs QWs underline the importance of cubic Dresselhaus SOI,
and provide evidence of a significant contribution from biaxial
strain. The negative sign of $\gamma$ agrees with theoretical work,
and is compatible with various experimental results for
$\alpha/\beta$, assuming $\alpha$ to be positive for electrons that
are pushed towards the substrate by the structure inversion
asymmetry of the quantum well.

\section{Acknowledgements}
We thank  M. Cardona, J. A. Folk and S. D. Ganichev for discussions
and would like to acknowledge financial supported from the KTI and
the NCCR Nano.
\bibliographystyle{prsty}
%\bibliography{spin}

\begin{thebibliography}{10}

\bibitem{Datta1990}
S. Datta and B. Das, Appl. Phys. Lett. {\bf 56},  665  (1990).

\bibitem{Loss1998}
D. Loss and D.~P. DiVincenzo, Phys. Rev. A {\bf 57},  120  (1998).

\bibitem{Wolf2001}
S.~A. Wolf, D.~D. Awschalom, R.~A. Buhrman, J.~M. Daughton, S. von
Molnar,
  M.~L. Roukes, A.~Y. Chtchelkanova, and D.~M. Treger, Science {\bf 294},  1488
   (2001).

\bibitem{Bychkov1984}
Y.~A. Bychkov and E.~I. Rashba, J. Phys. C {\bf 17},  6039  (1984).

\bibitem{Dresselhaus1955}
G. Dresselhaus, Phys. Rev. {\bf 100},  580  (1955).

\bibitem{Dyakonov1986}
M.~I. Dyakonov and V.~Y. Kachorovskii, Sov. Phys. Semicond. {\bf
20},  110
  (1986).

\bibitem{Jusserand1992}
B. Jusserand, D. Richards, H. Peric, and B. Etienne, Phys. Rev.
Lett. {\bf 69},
   848  (1992).

\bibitem{Dresselhaus1992}
P.~D. Dresselhaus, C.~M.~A. Papavassiliou, R.~G. Wheeler, and R.~N.
Sacks,
  Phys. Rev. Lett. {\bf 68},  106  (1992).

\bibitem{Miller2003}
J.~B. Miller, D.~M. Zumb\"uhl, C.~M. Marcus, Y.~B. Lyanda-Geller, D.
  Goldhaber-Gordon, K. Campman, and A.~C. Gossard, Phys. Rev. Lett. {\bf 90},
  076807  (2003).

\bibitem{Krich2007}
J.~J. Krich and B.~I. Halperin, Phys. Rev. Lett. {\bf 98},  226802
(2007).

\bibitem{Marushchak1983}
V.~A. Maruschak, M.~N. Stepanova, and A.~N. Titkov, Sov. Phys. Solid
State {\bf
  25},  2035  (1983).

\bibitem{Koralek2009}
J.~D. Koralek, C.~P. Weber, J. Orenstein, B.~A. Bernevig, S.-C.
Zhang, S. Mack,
  and D.~D. Awschalom, Nature {\bf 458},  610  (2009).

\bibitem{Kalevich1990}
V.~K. Kalevich and V.~L. Korenev, JETP Lett. {\bf 52},  230  (1990).

\bibitem{Kato2004}
Y. Kato, R.~C. Myers, A.~C. Gossard, and D.~D. Awschalom, Nature
{\bf 427},  50
   (2004).

\bibitem{Meier2007}
L. Meier, G. Salis, I. Shorubalko, E. Gini, S. Sch\"on, and K.
Ensslin, Nat.
  Phys. {\bf 3},  650  (2007).

\bibitem{Wilamowski2007}
Z. Wilamowski, H. Malissa, F. Sch\"affler, and W. Jantsch, Phys.
Rev. Lett.
  {\bf 98},  187203  (2007).

\bibitem{Studer2009}
M. Studer, S. Sch\"{o}n, K. Ensslin, and G. Salis, Phys. Rev. B {\bf
79},
  045302  (2009).

\bibitem{Chernyshov2009}
A. Chernyshov, M. Overby, X. Liu, J.~K. Furdyna, Y. Lyanda-Geller,
and L.~P.
  Rokhinson, Nat. Phys. {\bf 5},  656  (2009).

\bibitem{Norman2010}
B.~M. Norman, C.~J. Trowbridge, J. Stephens, A.~C. Gossard, D.~D.
Awschalom,
  and V. Sih, Phys. Rev. B {\bf 82},  081304  (2010).

\bibitem{Bernevig2005}
B.~A. Bernevig and S.-C. Zhang, Phys. Rev. B {\bf 72},  115204
(2005).

\bibitem{Shaw1981}
D.~W. Shaw, J. Electrochem. Soc. {\bf 128},  874  (1981).

\bibitem{Adachi1983}
S. Adachi and K. Oe, J. Electrochem. Soc. {\bf 130},  2427  (1983).

\bibitem{Cardona1986}
M. Cardona, N.~E. Christensen, M. Dobrowolska, J.~K. Furdyna, and S.
Rodriguez,
  Solid State Commun. {\bf 60},  17   (1986).

\bibitem{Cardona1988}
M. Cardona, N.~E. Christensen, and G. Fasol, Phys. Rev. B {\bf 38},
1806
  (1988).

\bibitem{Meier1984}
F. Meier and B. Zakharchenya, {\em Optical Orientation}
(North-Holland, New
  York, 1984).

\bibitem{Studenikin2003}
S. Studenikin, P. Coleridge, P. Poole, and A. Sachrajda, JETP
Letters {\bf 77},
   311  (2003).

\bibitem{Weng2004}
M.~Q. Weng, M.~W. Wu, and L. Jiang, Phys. Rev. B {\bf 69},  245320
(2004).

\bibitem{Jusserand1995}
B. Jusserand, D. Richards, G. Allan, C. Priester, and B. Etienne,
Phys. Rev. B
  {\bf 51},  4707  (1995).

\bibitem{Leyland2007}
W.~J.~H. Leyland, R.~T. Harley, M. Henini, A.~J. Shields, I. Farrer,
and D.~A.
  Ritchie, Phys. Rev. B {\bf 76},  195305  (2007).

\bibitem{Eldridge2010B}
P.~S. {Eldridge}, J. {H{\"u}bner}, S. {Oertel}, R.~T. {Harley}, M.
{Henini},
  and M. {Oestreich}, Condmat {\bf arXiv:1010.2142},    (2010).

\bibitem{Riechert1984}
H. Riechert, S.~F. Alvarado, A.~N. Titkov, and V.~I. Safarov, Phys.
Rev. Lett.
  {\bf 52},  2297  (1984).

\bibitem{Studer2009PRL}
M. Studer, G. Salis, K. Ensslin, D.~C. Driscoll, and A.~C. Gossard,
Phys. Rev.
  Lett. {\bf 103},  027201  (2009).

\bibitem{Frolov2009}
S.~M. Frolov, S. L\"uscher, W. Yu, Y. Ren, J.~A. Folk, and W.
Wegscheider,
  Nature {\bf 458},  868  (2009).

\bibitem{Larionov2008}
A.~V. Larionov and L.~E. Golub, Phys. Rev. B {\bf 78},  033302
(2008).

\end{thebibliography}

\end{document}